\documentclass[intlimits,twoside,a4paper]{article}
\usepackage[cp1251]{inputenc}
\usepackage{color}        
\usepackage[eqsecnum]{cmpj3}

\usepackage{bm}
\issue{2020}{23}{3}{33701}
\doinumber{10.5488/CMP.23.33701}

\title[Total energy calculations: metallic Zn QD's]%
{Total energy calculation for the metallic hcp phase of Zn in the bulk, layered, and quantum dot limits 
}
\author[D. Olgu\'in ]%
 {D.  Olgu\'in\footnote{
	Permanent address:
Departamento de F\'\i sica, Centro de Investigaci\'on
y de Estudios Avanzados del Instituto Polit\'ecnico Nacional,
C.P. 07300, Ciudad de M\'exico, M\'exico}
        }
\address{
 Centro de Investigaci\'on
y de Estudios Avanzados del Instituto Polit\'ecnico
Nacional--Unidad Quer\'etaro, Libramiento Norponiente
No. 2000, Fracc. Real de Juriquilla 76230, Santiago de Quer\'etaro,
Quer\'etaro, M\'exico
}
\date{Received November 12, 2020, in final form February 12, 2020}

\begin{document}

\maketitle

\begin{abstract}

The structural and electronic properties of the 
metallic hcp phase of Zn in 
the bulk, monolayer, bilayer, and quantum dot limits
have been studied 
by using total energy calculations.
From our calculated density of states  
and electronic band structure, 
in agreement with previous work, 
bulk hybridization of the Zn--$4s$, $3p$,  and $3d$ orbitals
is obtained. 
Furthermore, we found that this orbital hybridization 
is also obtained 
for the monolayer, bilayer, and 
quantum dot systems.
At the same time, we found 
that the Zn monolayer and bilayer systems
show electronic properties 
characteristic of 
lamellar systems, 
while the quantum dot system
shows the behavior predicted for a 0D system.

\keywords ab initio calculations, metallic Zn, quantum dots, 2D systems

\end{abstract}

\section{Introduction}

The study of the electronic properties of quantum dots (QDs) 
is very interesting for both basic and applied physics.
Small enough quantum dots can contain a few atoms, 
and their energy spectra are expected to behave 
like those of ``big artificial atoms'', showing discrete energy levels 
\cite{Kastner}.
To the best of our knowledge, 
the electronic and optical properties 
of semiconductor quantum dots  
have been summarized in early and recent interesting reviews found 
in the literature 
\cite{Lehtonen,Boxberg,Zwanenburg,Reimann}.
By contrast, for metallic quantum dots, a lack of work is evident.
However, important reviews, mainly on metallic clusters, 
can be found in~\cite{Brack,Cheshnovsky,deHeer,Furche,Haberlen,Taylor}.

Metallic quantum dots, in their nanocomposite form,
have been proposed for use in memory devices~\cite{CWLin,ZJin} and in coloring glasses.
By using the multicolor detection properties of
semiconductors, optical devices 
applicable in nanoscience and nanotechnology,
as well as in chemistry, biology, and medicine, 
can be developed~\cite{JZhang-IJHEner,Kang,%
Hatef,JZhang-JNRese,Olejnik,Sheng,Drazic-trsporte,Shahid,Dai,Haughn,%
Frade}.
As far as we know, metallic Zn quantum dots have not been studied or reported
in the recent scientific literature.
However, 
it is found that  
Zn  clusters were studied recently, 
where the reports show 
stable Zn clusters having different
stacking arrangement of the atoms 
as well as interesting electronic properties \cite{Piotrowski,%
JWang,KIokibe,GLGutsev,AAguado2018,AAguado2015}.
On the other hand, 
there is an increasing interest in different Zn-based nanoparticles, 
the same for biomaterials,
as well as in solid state physics, and other technological applications
\cite{AAli,YSuI,YSuII,Darroudi,Seyed,Cerovic}.
Therein, it has been showed that 
by using controlled conditions 
several Zn-based nanocrystals, passivated or dopped,
can be grown in the symmetric zinc-blende phase~\cite{SPal,Karanikolos,Goswani2006,%
Goswani2007,PChristian}.
It is also interesting to note   the report stating that 
at low temperature the 
system ZrZn$_2$ shows 
ferromagnetic and superconductor 
properties~\cite{Day,Pfleider}.

Although, no reports on Zn metallic nanocrystals
are known, recent work on Zn nanodisks 
\cite{Lin-disk,Devan,Lin-spher,Quique} 
showed that Zn metallic nanodisks, 
with sizes of 520 nm wide and 144 nm thick, 
can exhibit a mixture of metallic and semiconductor
properties~\cite{Lin-disk}.
Thus, with the aim of contributing to the study of the electronic
properties of metallic quantum dots, in this work, 
we present the study of the electronic band structure  
of metallic Zn quantum dots. 
For completeness, calculations for the bulk
and layered systems (monolayer and bilayer cases) 
are also presented.  
To show the stability of
the systems studied, 
we have calculated the cohesive energy, and 
our found values will be presented below.

Much of the theoretical work on modelling the
quantum dots has
been done under the assumption that the finite
thickness of a quantum dot, usually much smaller than
the lateral extension of the confinement,
can be neglected.
However, the 
typical size of a 
quantum dot ranges from a few Angstroms to micrometers~\cite{Kastner,Reimann}. 
A complete analysis of the electronic properties
of QDs 
requires the study of their behavior 
as a function of the QD size.
However, these additional calculations,
which require considerable
computational resources, are outside the scope of the present work.

The remainder of this paper is organized as follows. Section~\ref{modelo}
describes the 
computational strategy used in our calculations.
In section~\ref{resultados}, 
a discussion of our results on the calculated 
cohesive energies and electronic 
properties for the systems studied is presented. 
Finally, section~\ref{concluye} presents our conclusions.
\vspace{-0.5cm}
\section{Calculation method}~\label{modelo}
In our calculations, we used the full-potential linearized
augmented plane-wave method (FP-LAPW) as implemented in the WIEN2K 
code~\cite{Blaha}.
In this method, wave
functions, charge density, and potential are expanded in
spherical harmonics within non-overlapping muffin-tin
spheres, while plane waves are used in the remaining interstitial
region of the unit cell. In the code, the core 
states are treated with a
multiconfiguration relativistic Dirac-Fock approach, whereas
valence states are treated in a scalar relativistic approach.
The exchange-correlation energy is calculated using the
GGA correction of Perdew { et al.} \cite{Perdew,Perdew1}.
The atomic electronic configuration used 
was Zn [Ar] $4s$ $3d$; 
where the Zn $3p$ states were treated as
valence-band states using the local orbital extension of the
LAPW method \cite{Blaha}.

To find the minimum of the total energy for each system as 
a function of the variational parameters, the muffin-tin radius and 
the cutoff energy, 
a very careful step analysis was performed, 
where for the bulk case   
a value of 11.0 for $R_{\rm MT}$$K_{\rm max}$ 
(where $R_{\rm MT}$ is the muffin-tin radius and
$K_{\rm max}$ is the plane wave cutoff) and a  
value of 13.0 Ry for the cutoff energy parameter ($G_{\rm max}$) 
were used, while for the monolayer case the values used were
10.5 for $R_{\rm MT}$$K_{\rm max}$ 
and 13.0 Ry for $G_{\rm max}$, and for the 
bilayer case
the values used were $R_{\rm MT}$$K_{\rm max}$=10.5 
and $G_{\rm max}$=14.0 Ry.

At the same time, in our calculations, 
an appropriate set of {\bf k}-points 
in the irreducible sector of the Brillouin zone was used. 
Here, we used a value 
of 1600 k-points  for the bulk case, equivalent to a $(15\times 15\times 7)$
Monkhorst-Pack grid of the unit cell, 
and 2000 k-points for the layer systems, equivalent to 
$(20\times 20\times 4)$ Monkhorst-Pack grid 
\cite{Monk}.
It should be noted that for the quantum dots, 
the variational parameters used for
the total energy and electronic calculations 
were 
$R_{\rm MT}$$K_{\rm max}$=7.0,  
$G_{\rm max}$=12.0 Ry, and 
only the Gamma point of the first Brillioun zone
was used \cite{YXi-physicaE,Yamijala-CPLett}.

To study the Zn low dimensional systems, 
we  used the slab model. 
Therein, for the monolayer case,
one Zn-atomic layer
in the hcp lattice was constructed, 
while
for the bilayer case, we  used two Zn-atomic layers 
in the hcp lattice keeping the bulk hcp distance 
between the Zn layers, and 
for the QD we used a finite
slab containing thirteen atomic sites having 
obtained a pyramidal QD type,
whose dimensions were 
5.32 \AA\ for the basis and 2.65 \AA\ for the height.
Here, 
we  passivated the Zn-QD with Hydrogen atoms, since it had 
been showed that the passivation
stabilizes the bare QD \cite{Yamijala-CPLett}. 
In the inset of 
figures~\ref{dos-layer} and \ref{dos-nano}, 
the  slab models used are shown schematically, 
where we can distinguish 
the different atomic sites for metallic hcp Zn (red symbols).

To properly isolate each slab from their periodic images 
and to avoid large vacuum zones, 
which, in plane-wave methods, considerably increase the demand 
of computational resources, 
i.e., to keep the calculations accessible,
based on the fact that in bulk hcp zinc, 
the nearest neighbor distances 
range between 2.66 \AA\ and 2.91 \AA\ for the in-plane 
and interlayer neighbors, 
respectively, 
a vacuum of 10 \AA\ 
with the upper slab was considered for the layered systems,  
whereas a vacuum of 13 \AA\ for the upper side
and 8 \AA\ for the lateral side was used for the QD system. 
\vspace{-5mm}
\section{Results} 
\label{resultados}

Before a detailed discussion of the electronic properties 
of Zn systems,  we first show our calculated structural properties
of the bulk crystal.
The Zn hexagonal crystal structure  
is showed in the inset of figure~\ref{dos-bulk}. 
Our calculated lattice parameters are $a = 2.6712 $ 
\AA\ and
$ c = 4.9586 $ \AA. 
In excelent agreement with the experimental values:
$a= 2.6649 $ \AA\  and
$c = 4.9468$ \AA\ 
\cite{Zn-lattice}.

Then, for the layered systems we  relaxed the atomic positions in the slab 
using the Hellmann-Feynman forces until the force component 
was less than 0.5 mRy/au.
Our results show that,
for the monolayer case, the geometry is rigid and does not
allow displacements for the in-plane nearest neighbor distances, and 
for the bilayer case we found 
that the interlayer distances
slightly increase  from 2.93 \AA\ (the bulk value) to 2.98 \AA.
Meanwhile, for the QD case, the system relaxes from 2.93 \AA\ to 2.90 \AA,
and the in-plane nearest neighbor distances remain almost constant, where
the initial and final values are 2.66~\AA\ and 2.67~\AA, respectively.
In order to show the trend in the stability of the layered systems, 
we  calculated the cohesive energy\cite{DWang-cohesive} 
\[
E_{\rm coh} = 
\frac{N_{\rm Zn} E^{\rm free}_i -E^{\rm supercell}_{\rm Tot}}{N_{\rm Zn}}\,,
\]
where, 
$E_i^{\rm free}$ is the total energy of the free $i$-th atom,
$N_{\rm Zn}$ the number of Zn atoms in the layer system, 
and $E_{\rm Tot}^{\rm supercell}$
the total energy of the supercell.
We found that the cohesive energy for the bilayer case is 0.73~eV/atom,
while for the monolayer case the cohesive energy is 0.56 eV/atom. 

To calculate the cohesive energy for 
the hydrogen passivated quantum dot, 
we need to compute the binding energy 
\[
E_{\rm binding} =N_{\rm Zn} E^{\rm free}_{\rm Zn} +
N_{\rm H} E^{\rm free}_{\rm H} -
E^{\rm QD}_{\rm Tot}, 
\]
and then, by removing  the Zn-H energy 
energy from the binding energy,
we will have the cohesive energy for the QD~\cite{Zdetsis-hydrogen, Zdetsis-hydrogen-ol}, 
i.e., 
\[
E^{\text{QD}}_{\rm coh} = \frac{E_{\rm binding} + \mu_{\text H} N_{\text H}}{N_{\rm Zn}}\,,
\]
where $\mu_{\text H}$ is the hydrogen chemical potential. 
From this expression, we find that  
the cohesive energy for the QD 
is 0.29 eV/atom.
Considering the relatively low values for the cohesive energies,
we found that our calculated systems could be considered stable with weak 
chemical bonds. Thus, from the experimental point of view,
special attention will be necessary to study these systems. 

\subsection{Electronic properties: Zn bulk}

\begin{figure}[!t]
\includegraphics[width=0.45\textwidth]{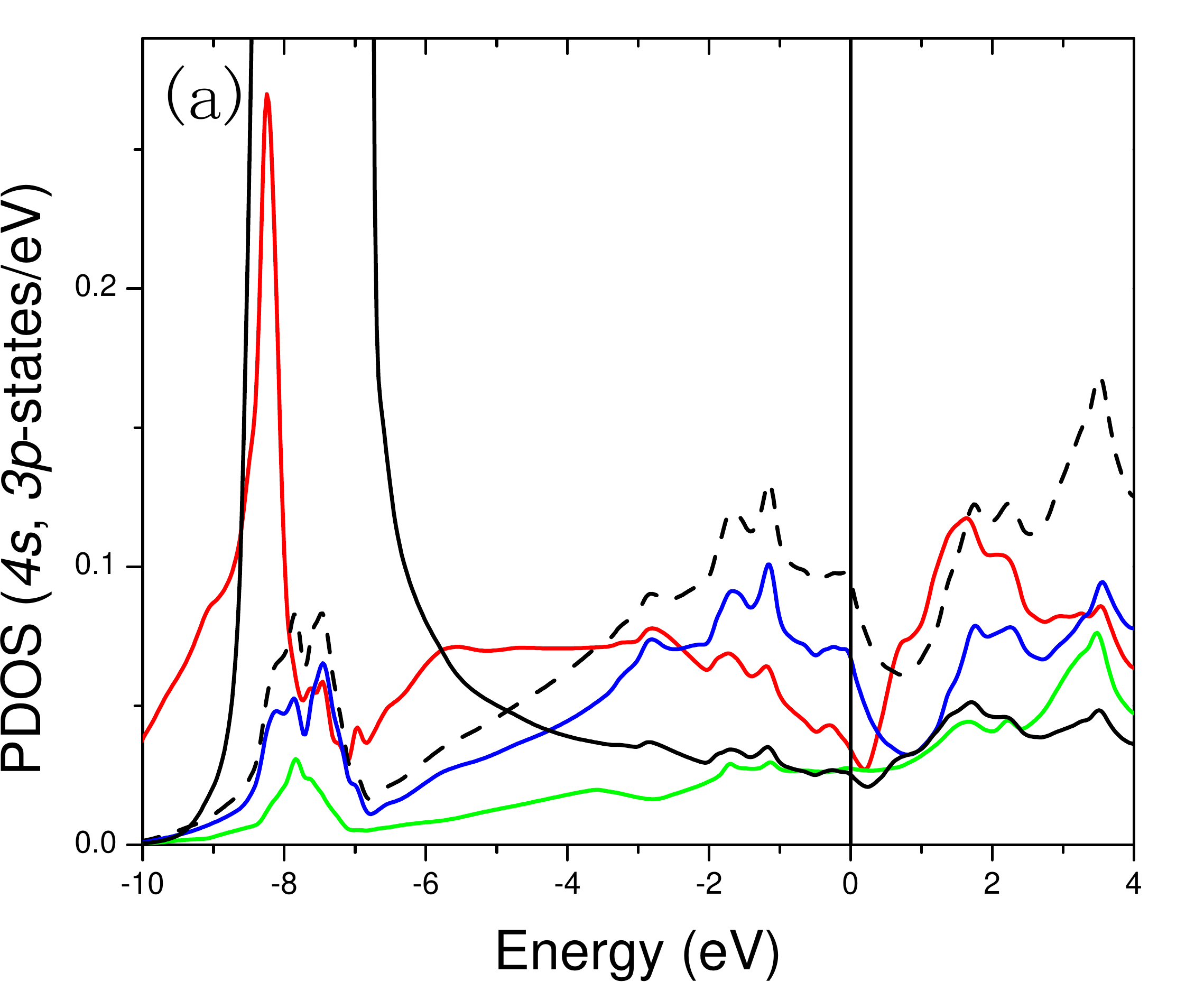}
\includegraphics[width=0.45\textwidth]{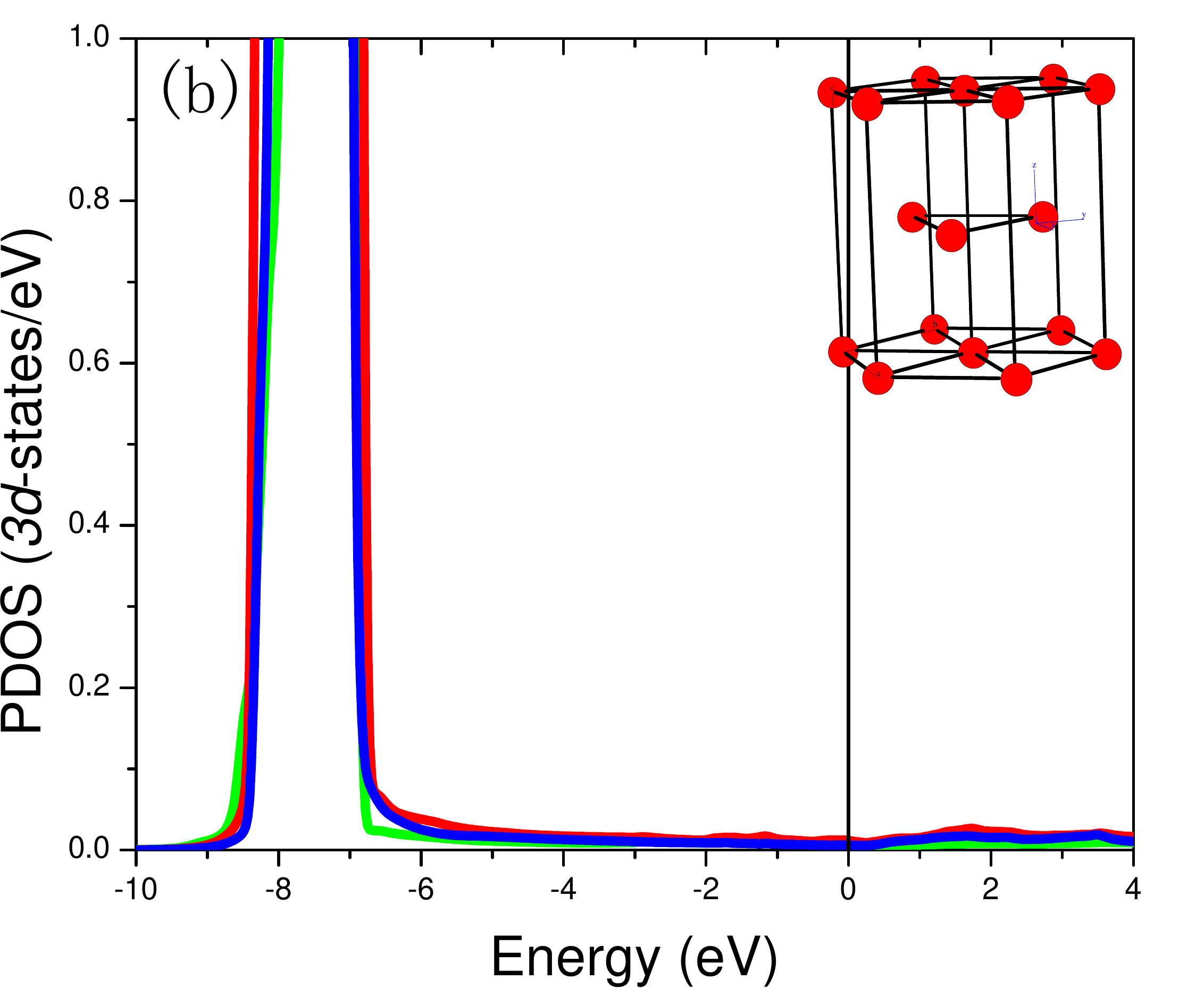}
\caption{(Color online) 
The calculated partial density of states (PDOS) for metallic hcp Zn. 
The left-hand
side panel shows the $4s$ (red lines), $3(p_x+p_y)\ ({\rm blue\ lines})$ and 
$3p_z\ ({\rm green\ lines})$ orbitals; 
for comparison, 
the $3p$ obitals (black dashed lines), and 
the $3d$ orbitals (black lines) are also shown.
The right-hand side panel shows the $3d_{z^2}$ (green lines),
$3(d_{x^2+y^2}+d_{xy})$ (red lines), and $3(d_{xz}+d_{yz})$ (blue lines) 
orbitals; the inset shows the hcp unit cell.
The zero of the energies represents the Fermi level.
}
\label{dos-bulk}
\end{figure}

Figure \ref{dos-bulk} shows the 
calculated partial density of states (PDOS)  
for the bulk hcp Zn. 
For the sake of clarity, 
the left-hand panel shows the Zn--$4s$ and $3p$ orbitals, 
while the right-hand panel shows the Zn--$3d$ orbitals.
The plot shows that 
there is a non-zero population of states at the Fermi level,
that is, we have obtained a system with metallic characteristics,
as expected.

From the plot, we can distinguish 
three energy regions where the calculated states are located:
below $E_{\text F}$, the first region with energies from $-10.5$ to $-6.7$ eV, 
where the main contribution to the PDOS
arises from the $3d$ orbitals; 
states that
are mainly located in the energy range from $-8.5$ eV to $-6.7$ eV;
a second range of energies 
from $-2.9$ eV to 0.0 eV; 
and a third energy region located above $E_{\text F}$,  
with energies ranging from 1.6 eV to 3.6 eV.

From the left-hand side 
panel and from the comparison 
of the different calculated states shown, 
we can observe the already known 
hybridization of the Zn--$4s$, $3p$,  and $3d$ 
orbitals, in good agreement 
with earlier work~\cite{Daniukt,Singh,Blaha-zn,Papa-zn}.

A detailed analysis of the calculated 
PDOS shows that,  in addition to the highly localized 
3$d_{z^2}$ (green lines),
the 3($d_{x^2+y^2}$+$d_{yx}$) (red lines), and the 
3($d_{xz}$+$d_{yz}$) orbitals (blue lines),
the bands whose peaks are located at $-8.25$ eV
have important contributions from the $4s$ orbitals.
The main peaks for the $3(p_x+p_y)$ orbitals (blue lines) 
are located in the energy ranges from $-8.13$ eV to $-7.45$~eV and 
from $-2.9$~eV to $0.0$~eV, as well as at energies above $E_{\text F}$, 
from $1.6$~eV to $3.6$~eV. 
The contributions of the $3p_z$ orbitals (green lines) 
are obtained at $-7.8$ eV and $3.5$~eV.

\begin{figure}[!t]
\includegraphics[width=0.33\textwidth]{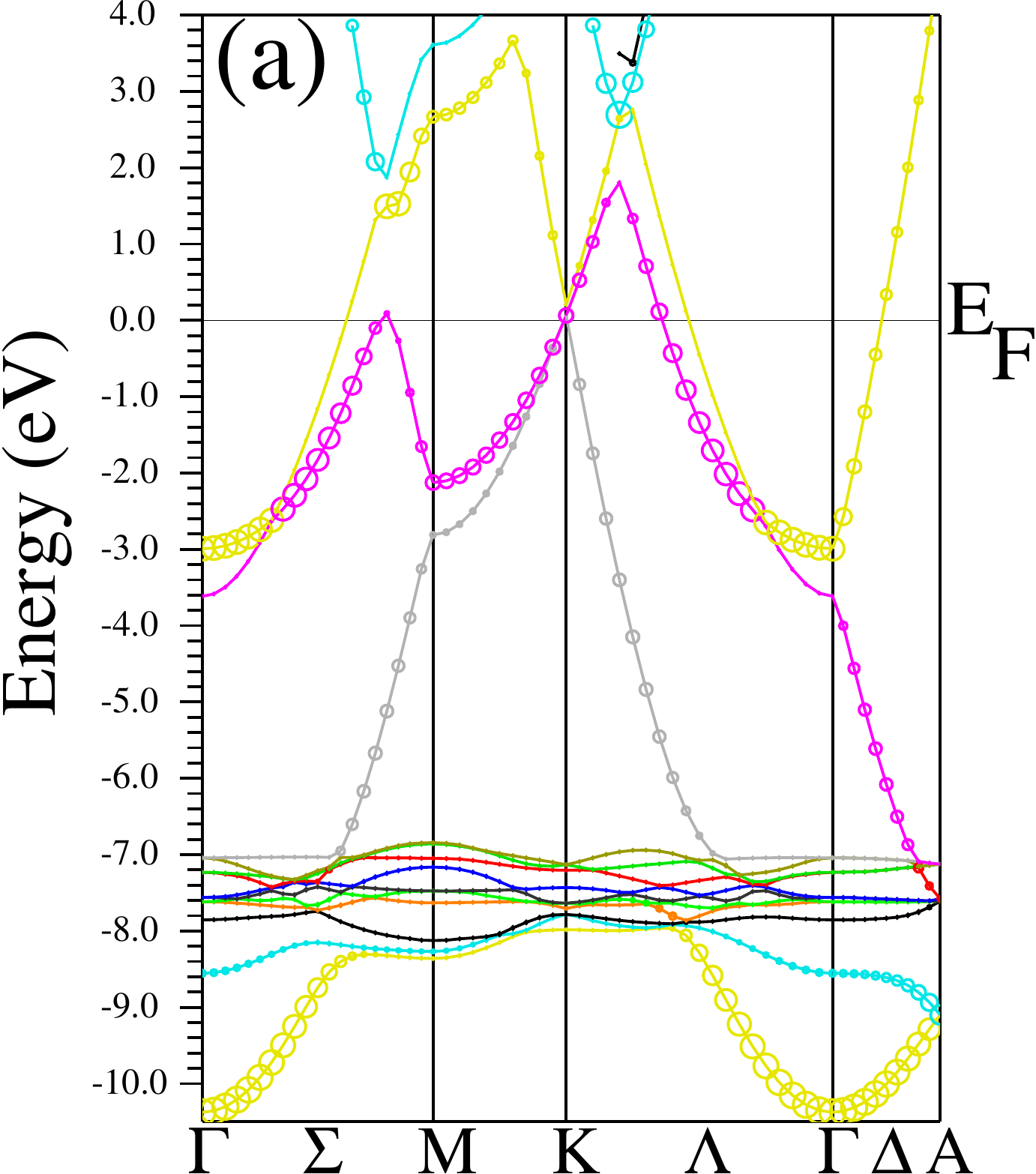}
\includegraphics[width=0.33\textwidth]{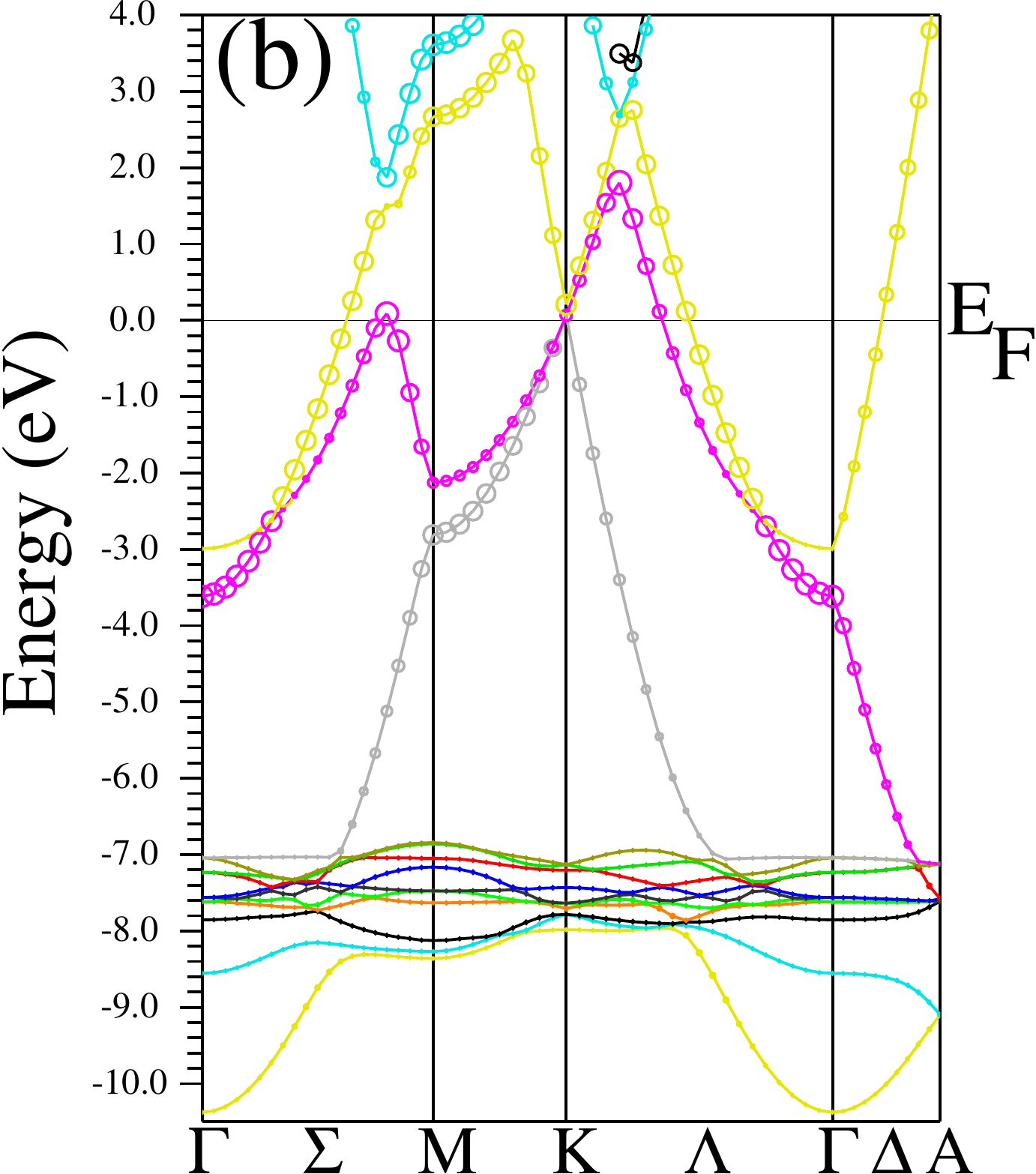}
\includegraphics[width=0.33\textwidth]{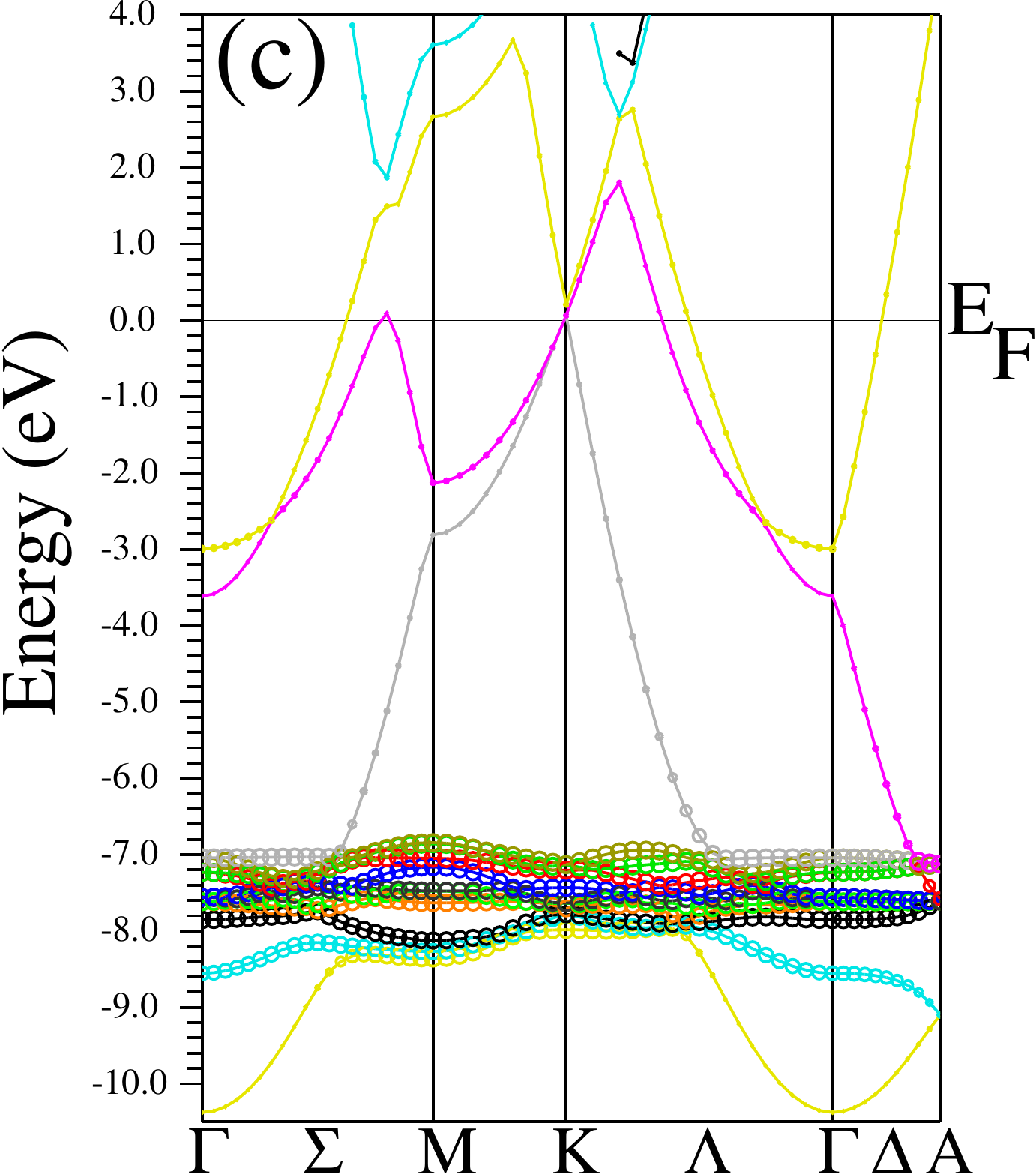}
\caption{(Color online) 
The calculated
atomic orbital contributions to the 
electronic band structure of 
bulk hcp Zn, where
the orbital character of the electronic bands
is represented by circles 
(colors are a guide to the eye).
The left-hand panel shows the $4s$ orbitals, the middle panel
shows the $3p$ orbitals, and the
right-hand panel shows the $3d$ orbitals.
}
\label{bnds-bulk}
\end{figure}

Further information on the calculated electronic states
can be obtained from the analysis of the atomic 
orbital contributions to the electronic band structure,
as shown in figure~\ref{bnds-bulk},
where the atomic orbital contributions 
to the electronic band structure 
are depicted by circles. 
A larger circle size
corresponds to a higher charge within the atomic sphere, 
qualitatively showing  
those orbitals that constitute each electronic band~\cite{Blaha}.

Figure~\ref{bnds-bulk}~(a) shows that 
the lower band located in the energy range from $-10.5$ eV
to $-8.0$ eV, showing high dispersion,  
originates from the $4s$ states (yellow circles) 
and hybridizes with the $3d$ orbitals.
Figure~\ref{bnds-bulk}~(a) and figure~\ref{bnds-bulk}~(b) show that 
in the energy range from --7.0 eV to 3.0 eV, 
the electronic bands are a hybridization of 
the $4s$ and $3p$ states (magenta and gray circles). 
Figure~\ref{bnds-bulk}~(c)
shows that the $3d$ orbitals are mainly localized 
in the energy range from $-8.5$ eV to $-6.7$ eV,
as we have commented above.

\subsection{Zn-monolayer case and Zn-bilayer case} 

\begin{figure}[!b]
\includegraphics[width=0.50\textwidth]{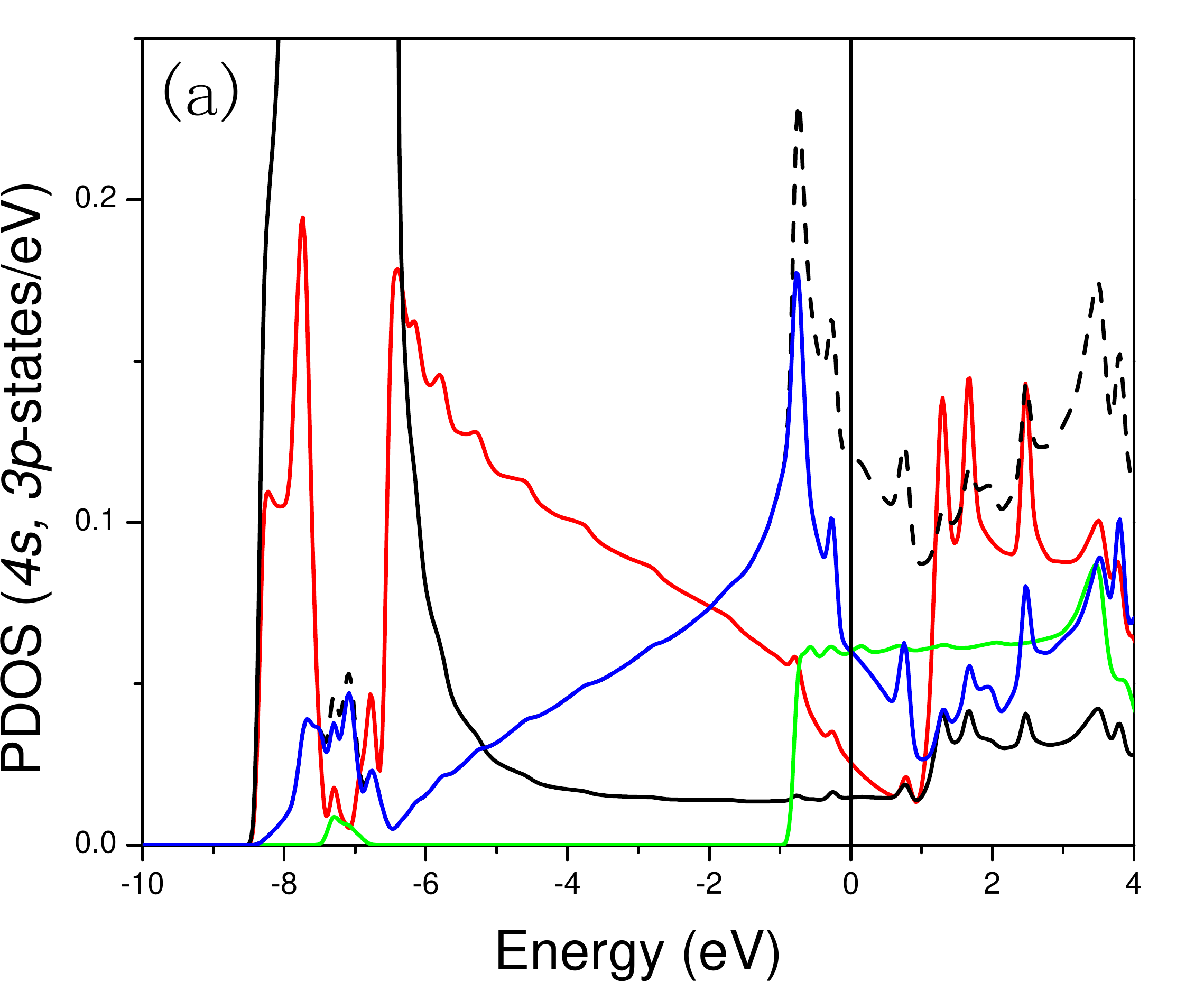}
\includegraphics[width=0.50\textwidth]{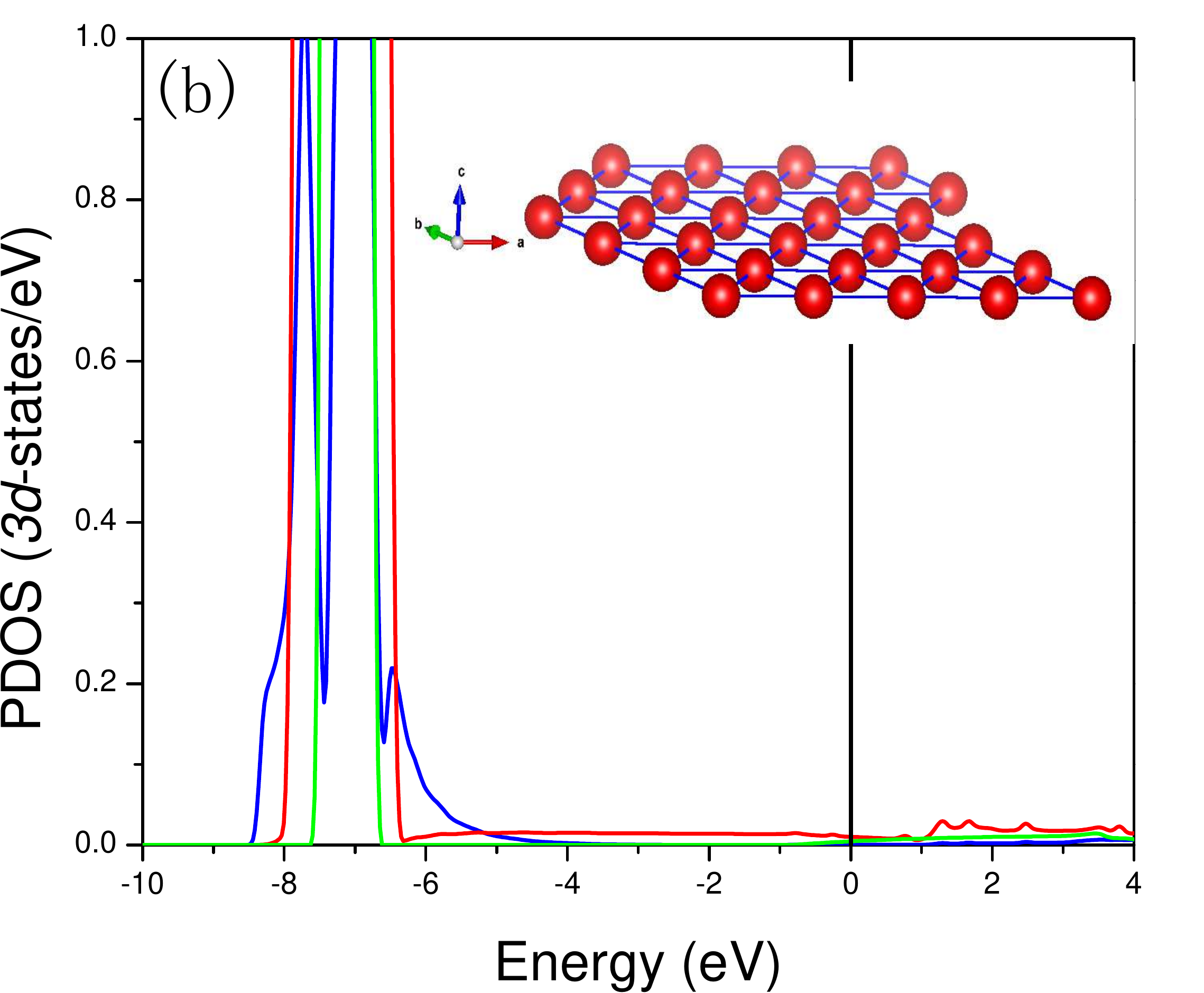}
\includegraphics[width=0.50\textwidth]{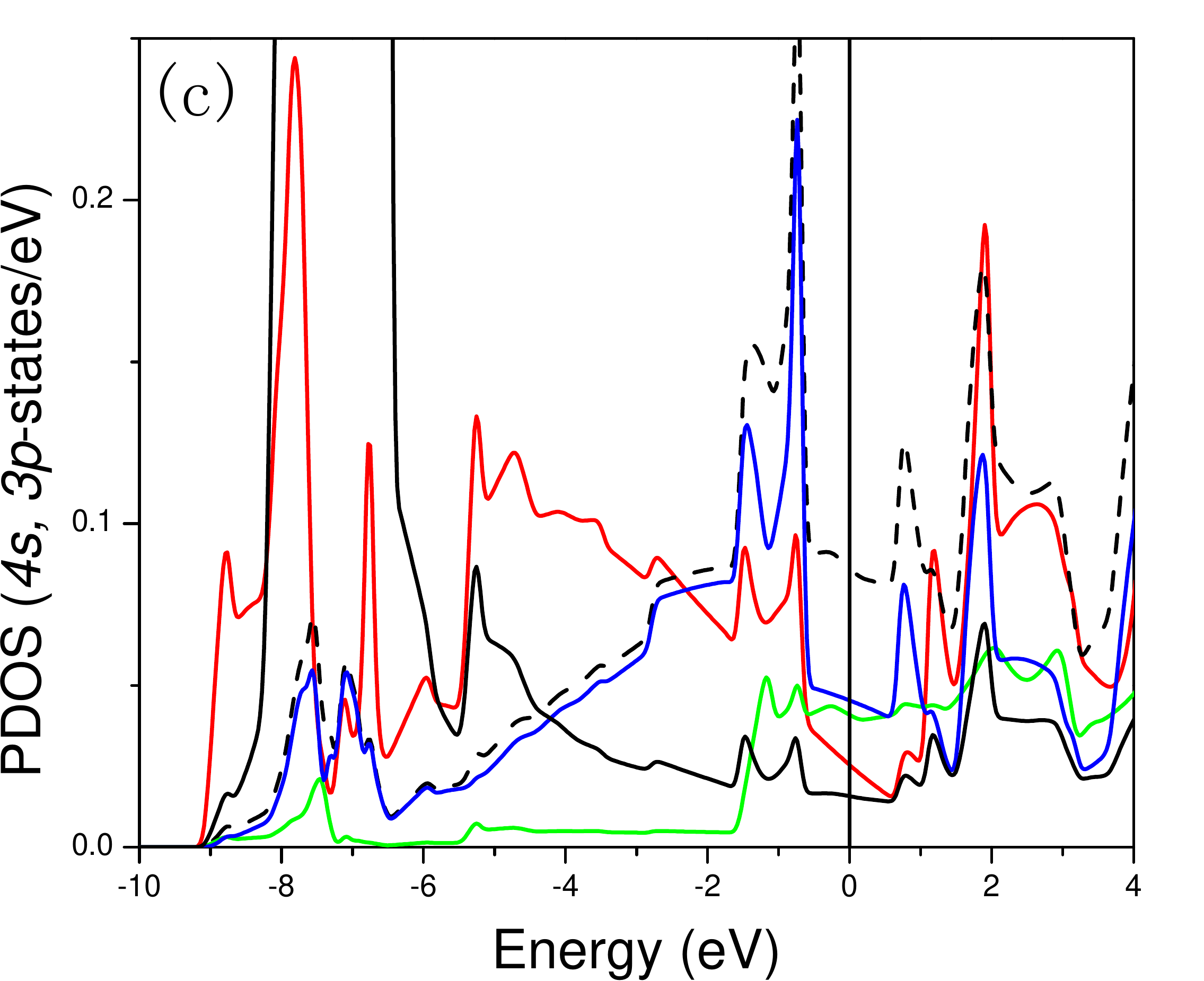}
\includegraphics[width=0.50\textwidth]{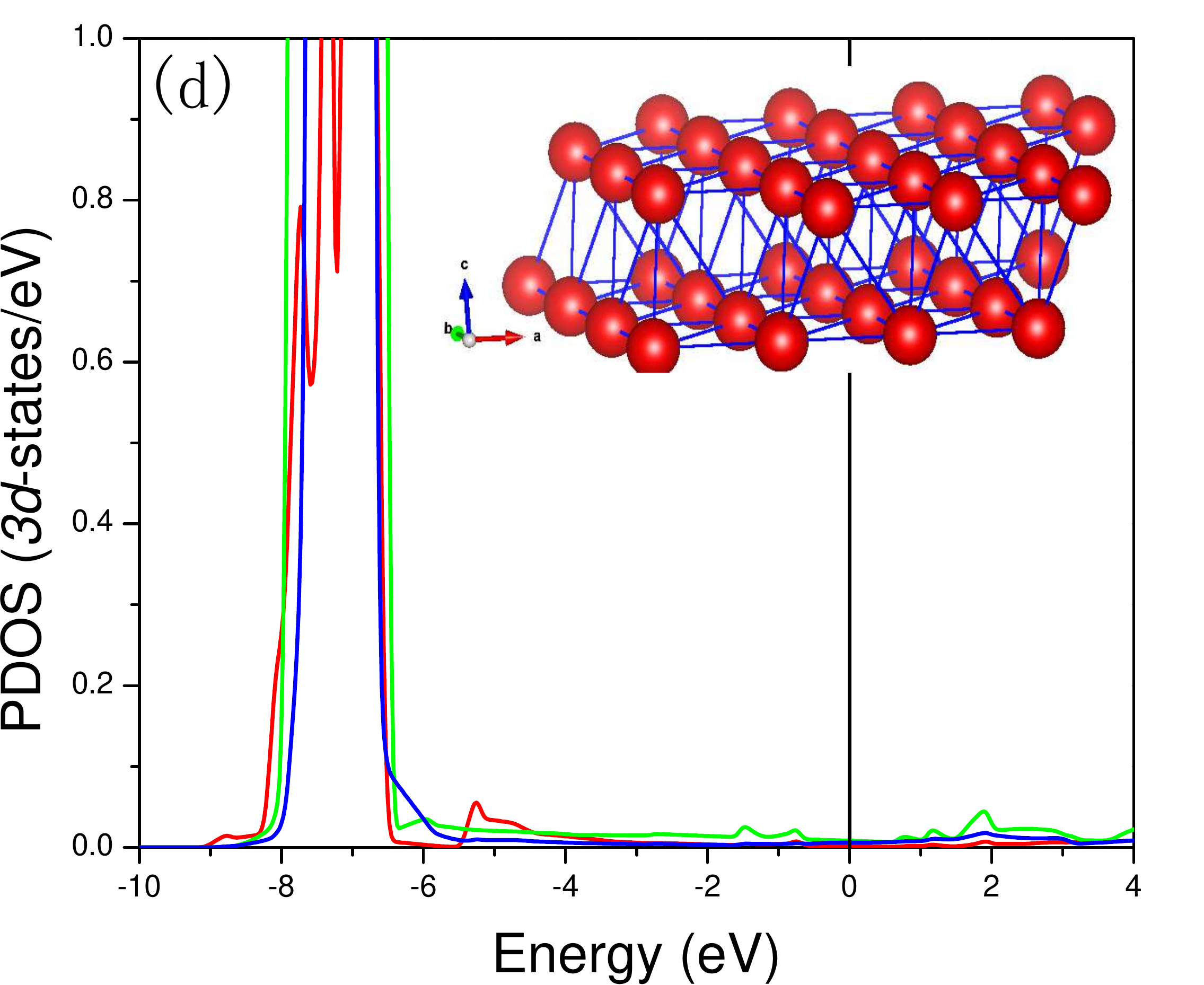}
\caption{(Color online) 
Same as figure~\ref{dos-bulk} but for
the Zn-monolayer case, upper panel,
and the Zn-bilayer case, lower panel. 
The inset in the right-hand panel shows a 
schematic of the slab model used in our
calculations.
}
\label{dos-layer}
\end{figure}

Figure~\ref{dos-layer} shows our 
calculated PDOS for the layered systems, 
where the upper panel depicts the
results for the  Zn-monolayer case  
and the lower panel depicts those for the 
Zn-bilayer case.
As in the bulk case, 
we see that 
both systems have metallic characteristics. 

At the same time, 
our results show 
three energy regions 
where the calculated states are located.
Two of these energy regions 
are below $E_{\text F}$, 
the lower one in the energy range from approximately $-9.2$ eV to $-6.8$ eV,
where the $3d$ states are located, 
and the second one in the energy range from $-6.4$ eV to $0.9$ eV, 
where the $4s$ and $3p$ states are located.
Above $E_{\text F}$, a third energy range from 0.9 eV to 3.4 eV 
is obtained, where  a second set of $4s$ and $3p$ states is located.

Using the bulk values as a reference, 
a detailed comparison of the calculated energies 
for the layered systems shows that 
the $3d$ states are located in the energy range from $-8.5$ ($-9.2$) eV
to $-6.2$ ($-6.8$) eV (where the first values refer to
the monolayer case and the values in parentheses correspond to the 
bilayer case),
almost at the same energy values already found for the bulk case. 

By contrast, the calculated values for 
the $4s$ and $3p$ states move up slightly in energy.
For example, the main peak of the lower $4s$ bands is now 
located at $-7.7$ ($-7.8$) eV.
Furthermore, a second peak for the 
$4s$ bands is obtained at $-6.8$ ($-6.8$)~eV; 
although this band is also present in the bulk case,
it is more noticeable in the layered cases.
Additionally, 
in the energy range from $-6.4$~eV to 0.6~eV, 
a new shape 
of the $4s$ orbitals is obtained;
here, these states change from a constant step-like form 
to a sawtooth form  
(compare with figure~\ref{dos-bulk}).
At energies above $E_{\text F}$ 
we obtain three peaks arising from the $4s$ states, 
one peak located at 1.3 (1.3)~eV, another 
located at 1.8 (1.8)~eV,
and a third located at 2.5~eV.
For comparison, note that in the bulk case,
only one band
is obtained, namely, the one located at 1.6 eV. 

\begin{figure}[!t]
\includegraphics[width=0.33\textwidth]{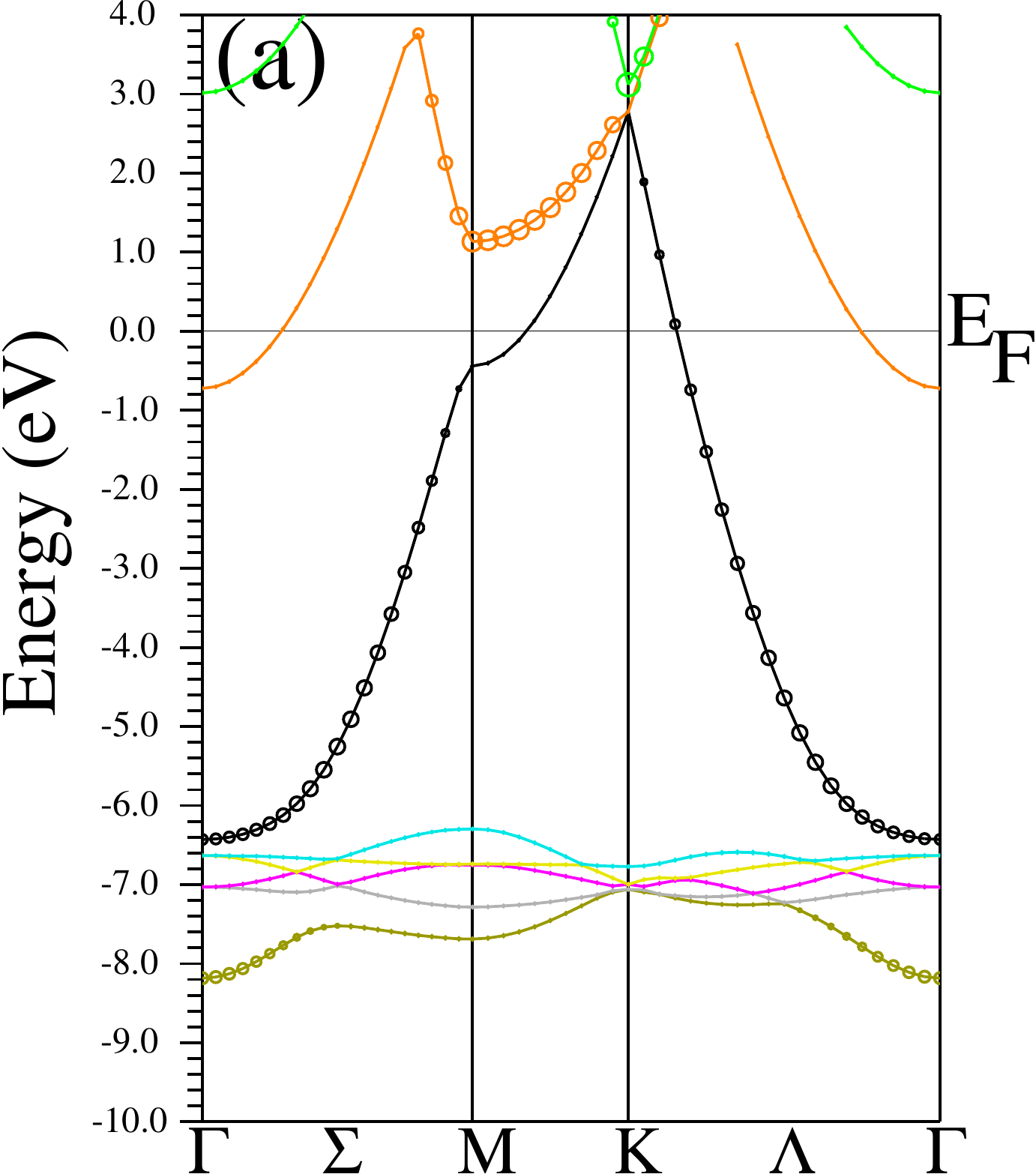}
\includegraphics[width=0.33\textwidth]{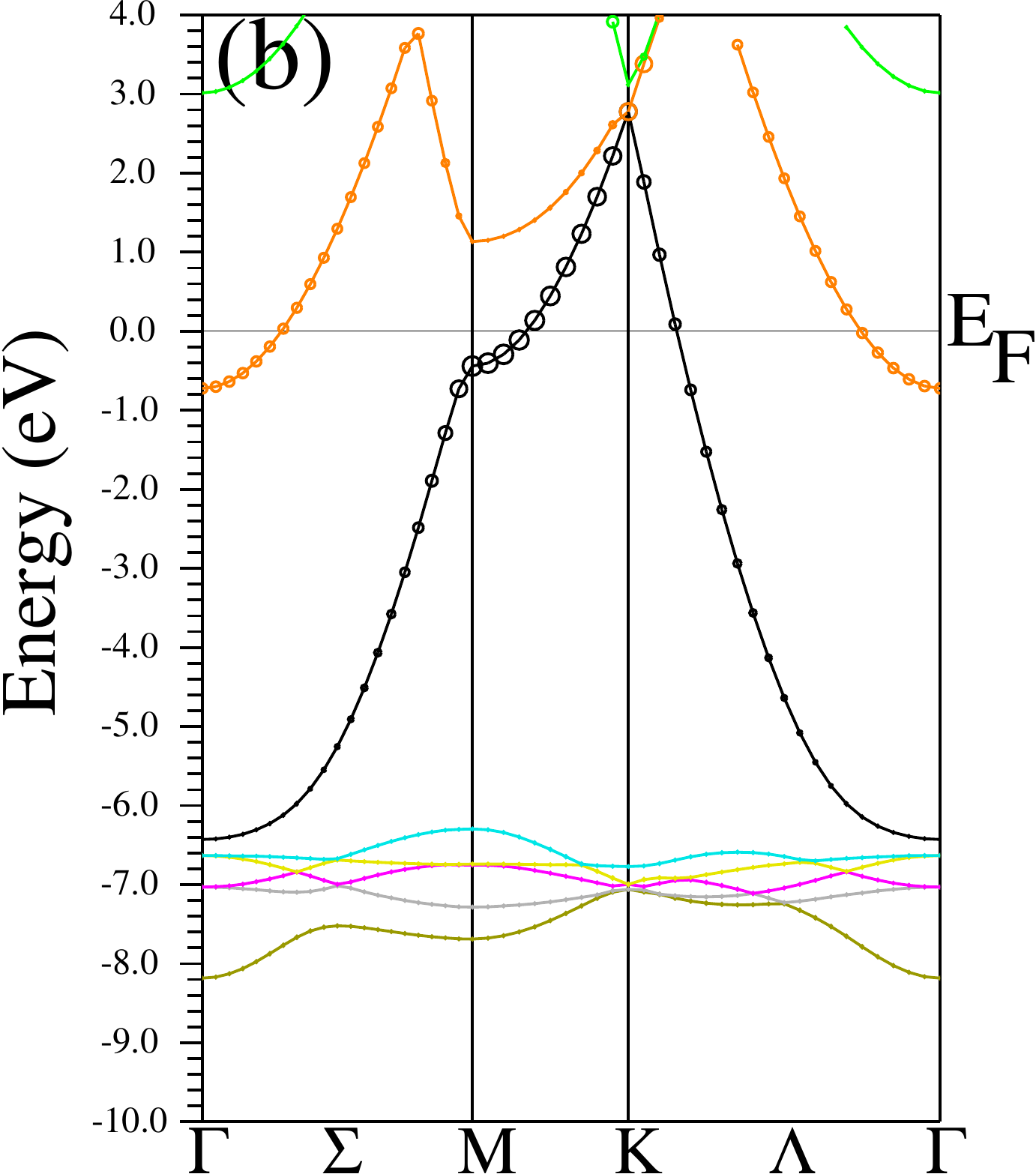}
\includegraphics[width=0.33\textwidth]{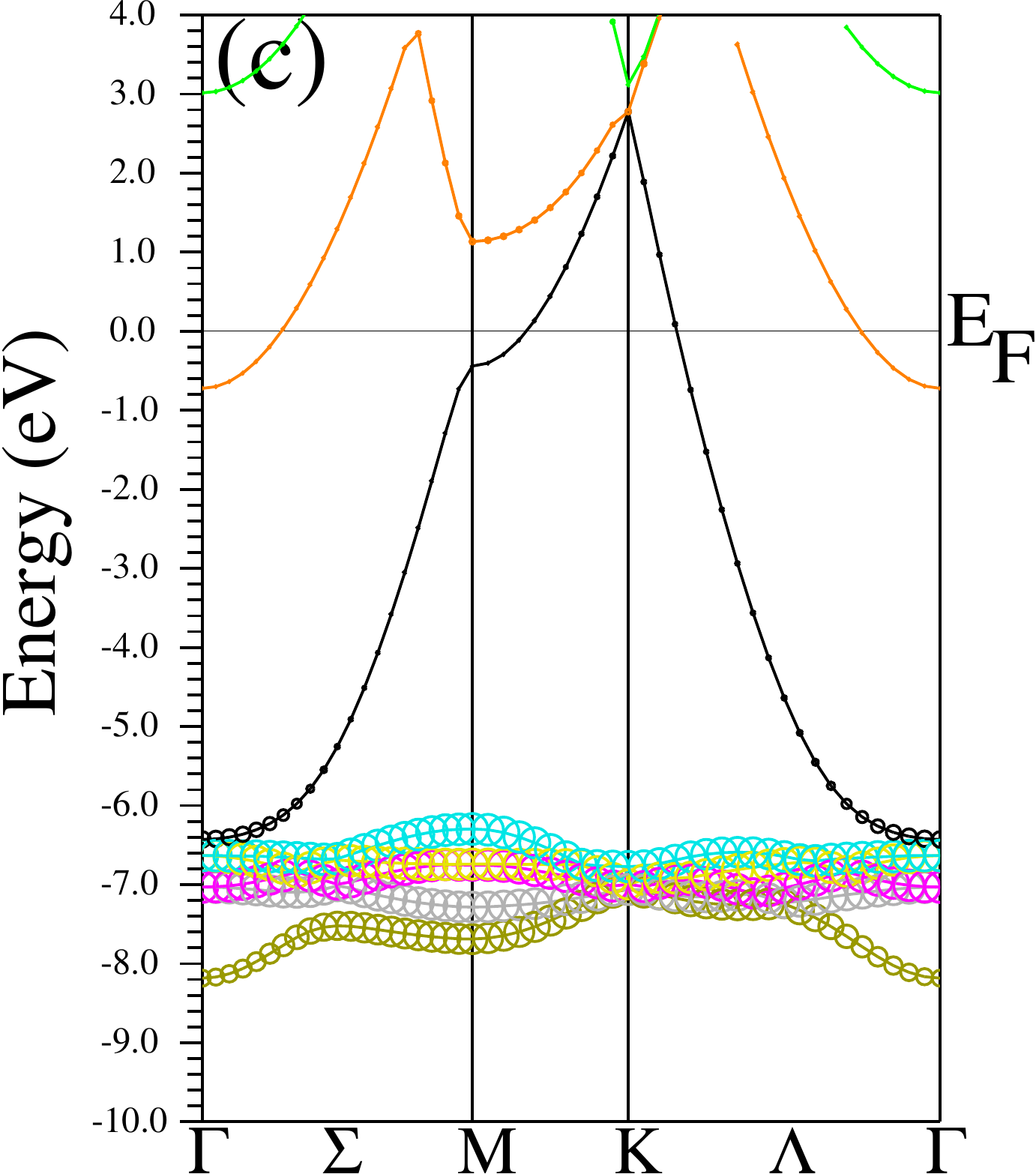}
\includegraphics[width=0.33\textwidth]{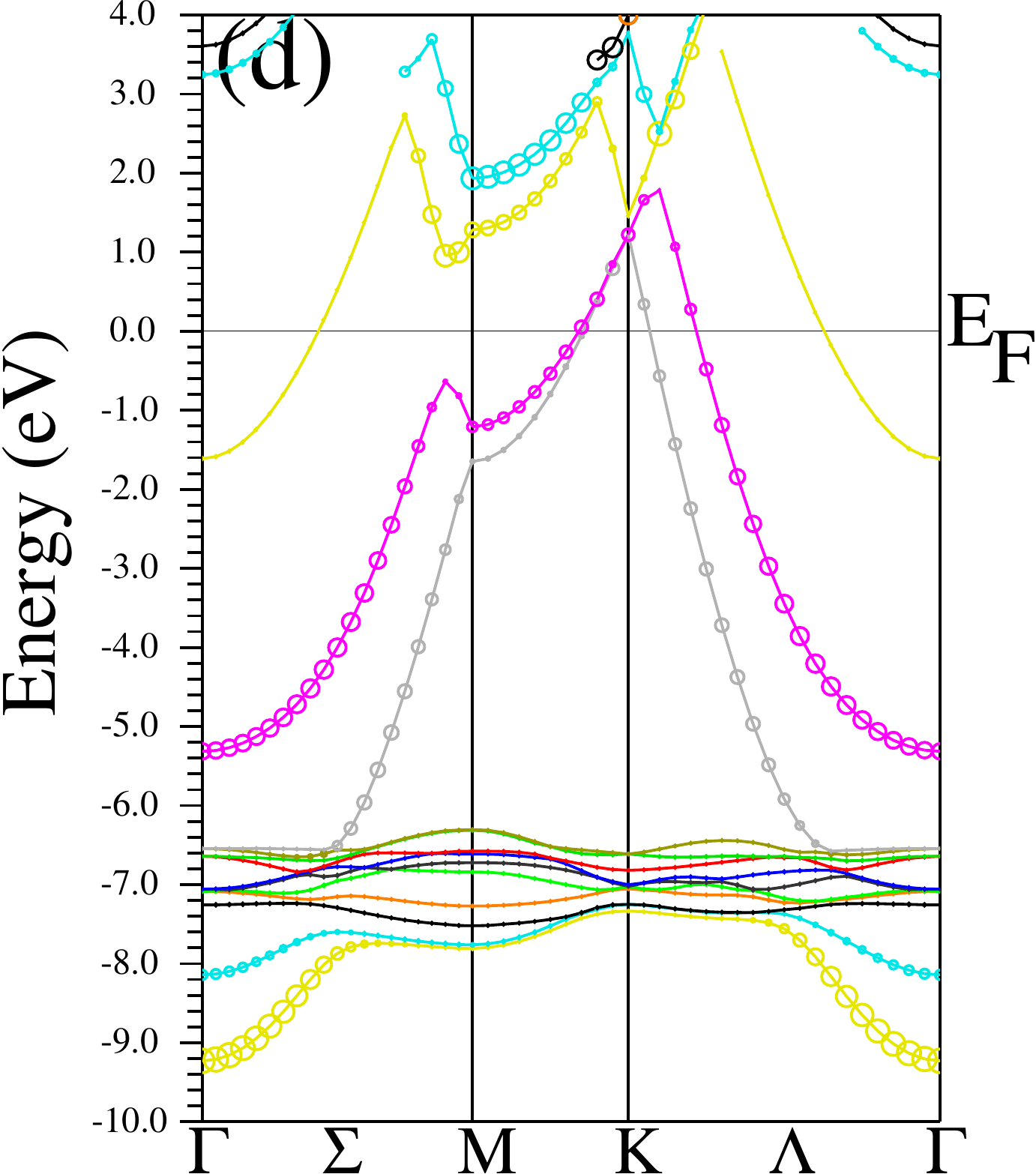}
\includegraphics[width=0.33\textwidth]{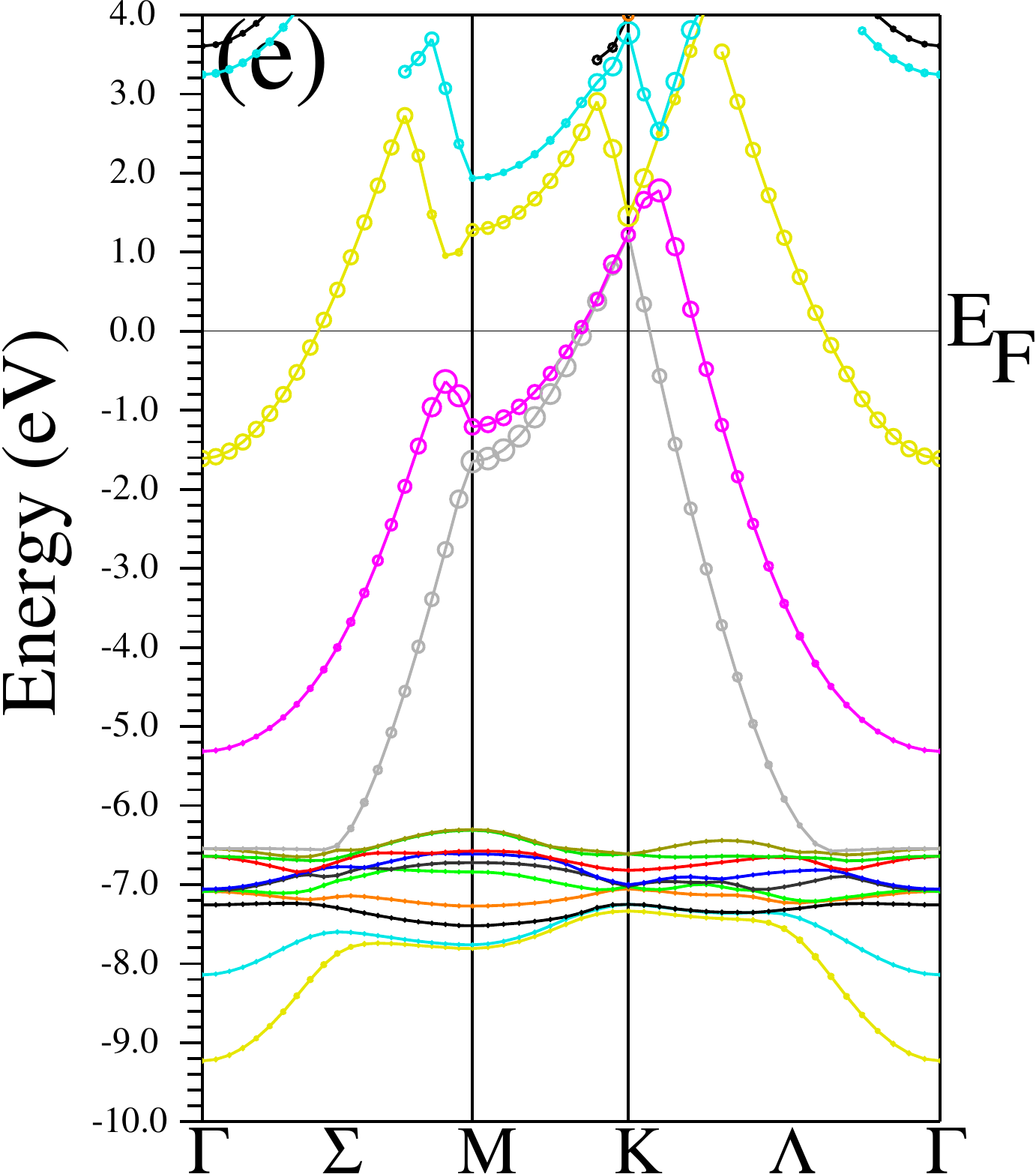}
\includegraphics[width=0.33\textwidth]{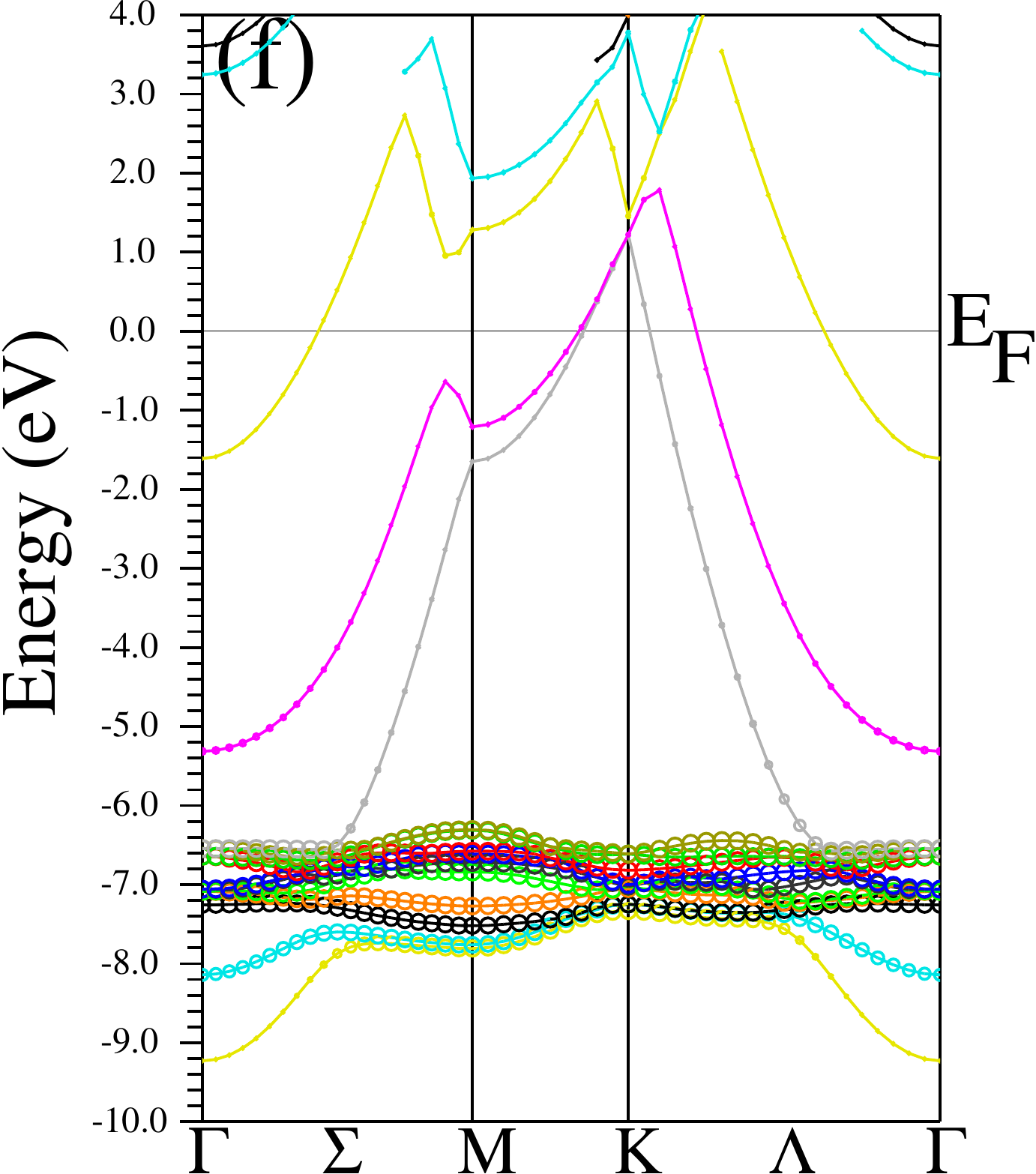}
\caption{(Color online) 
Same as figure~\ref{bnds-bulk} but for 
the Zn monolayer (upper panel) and bilayer (lower panel) cases. 
}
\label{bnds-half}
\end{figure}

The $3p$ orbitals also changed in shape compared with the bulk
case; here, the contribution of the $3(p_x+p_y)$ orbitals (blue lines)
to the PDOS in the energy range from $-7.7$ ($-7.6$)~eV to $-6.8$ ($-7.2$)~eV
seems to be remarkable. 
There are other noticeable peaks at $-0.8$ ($-1.6$)~eV
and $-0.3$ ($-0.8$)~eV along with the other peaks in the conduction band
at 0.7 (0.8) eV, 1.8 (1.8) eV, and 2.5 (1.8) eV.
The contributions of 
the $3p_z$ orbitals (green lines)
are obtained at the energy
range from $-1.5$ eV to 3.3 eV, 
showing a step-like shape at these energies.

The obtained changes in the PDOS
can also be observed from the calculated 
atomic orbital contributions to the 
electronic band structure, as shown in figure~\ref{bnds-half},
where, due to the different coordination 
in the apical direction 
between the layered and bulk cases, 
the number of bands obtained 
for the monolayer case is half that for the 
bilayer and bulk cases.

The calculated energy differences for the $4s$ and $3p$ states
can be summarized as follows: 
in the layered cases, 
the $4s$ states 
located in the energy range from $-8.5$ ($-9.3$)~eV to $-7.3$ ($-7.3$)~eV
show less dispersion than those in the bulk case.
Additionally,  
in the energy range from
$-6.4$ ($-7.0$)~eV to $0.0$~eV, 
the calculated energy values at the high symmetry points 
move up in energy for the layered systems 
(compare figure~\ref{bnds-bulk} and figure~\ref{bnds-half}).

Finally, we conclude that, 
except for the differences in the calculated energy values, 
in general, the calculated 
electronic properties of the layered systems  
show a pattern similar 
to that found for the bulk case,
revealing that the in-plane coordination produces 
lamellar systems, 
an interesting property of several
2D systems that 
are the object of recent and intensive work 
~\cite{Ullah-graph,Singh-graph,Kaloni-silicene,Rudenko,%
ShZhang-Arsenene-Antimonene,AngelRubio-selenene-tellurene,%
JZhou-ZnSe-layer}.

\subsection{Zn quantum dot}

\begin{figure}[!t]
\includegraphics[width=0.5\textwidth]{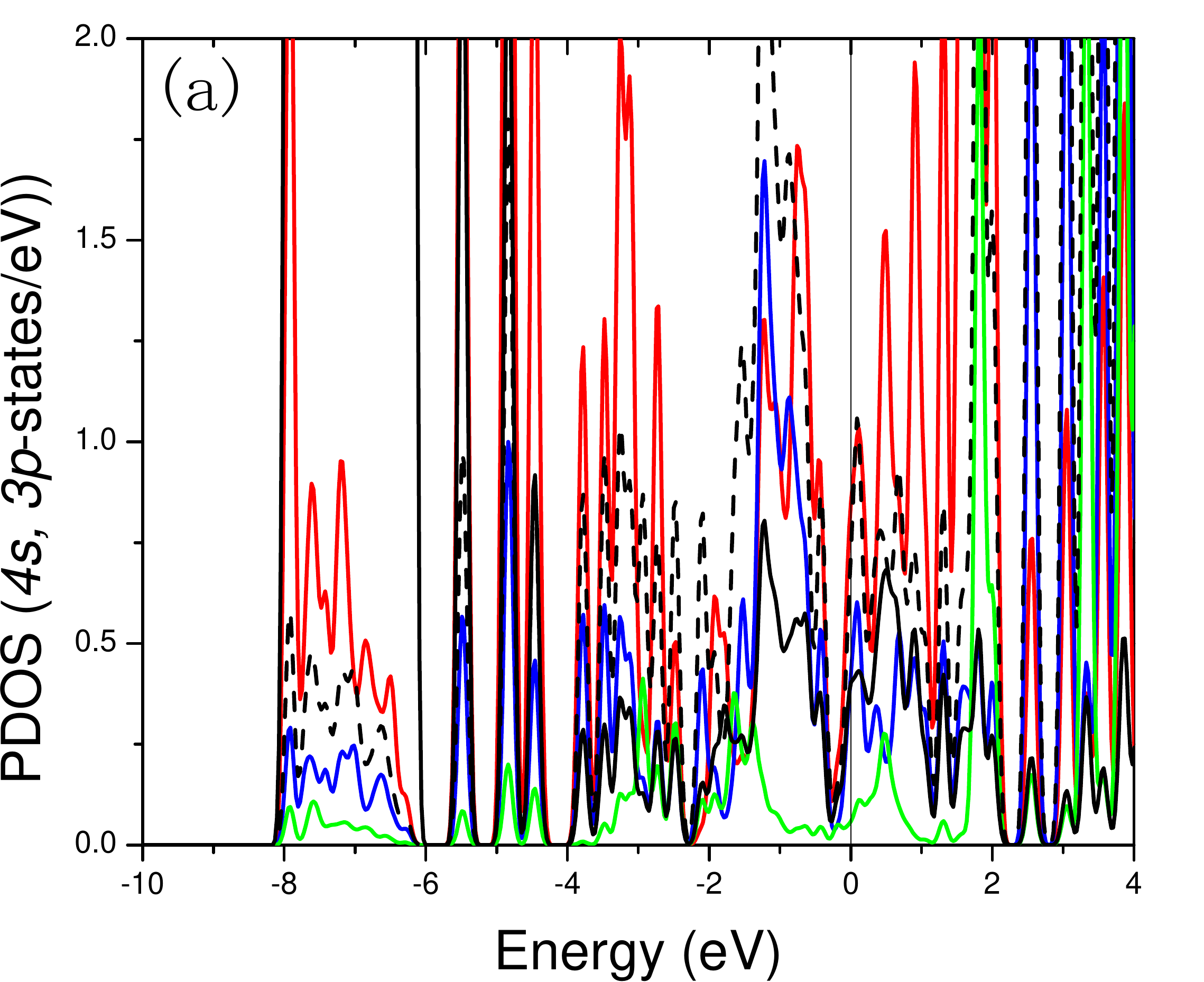}
\includegraphics[width=0.5\textwidth]{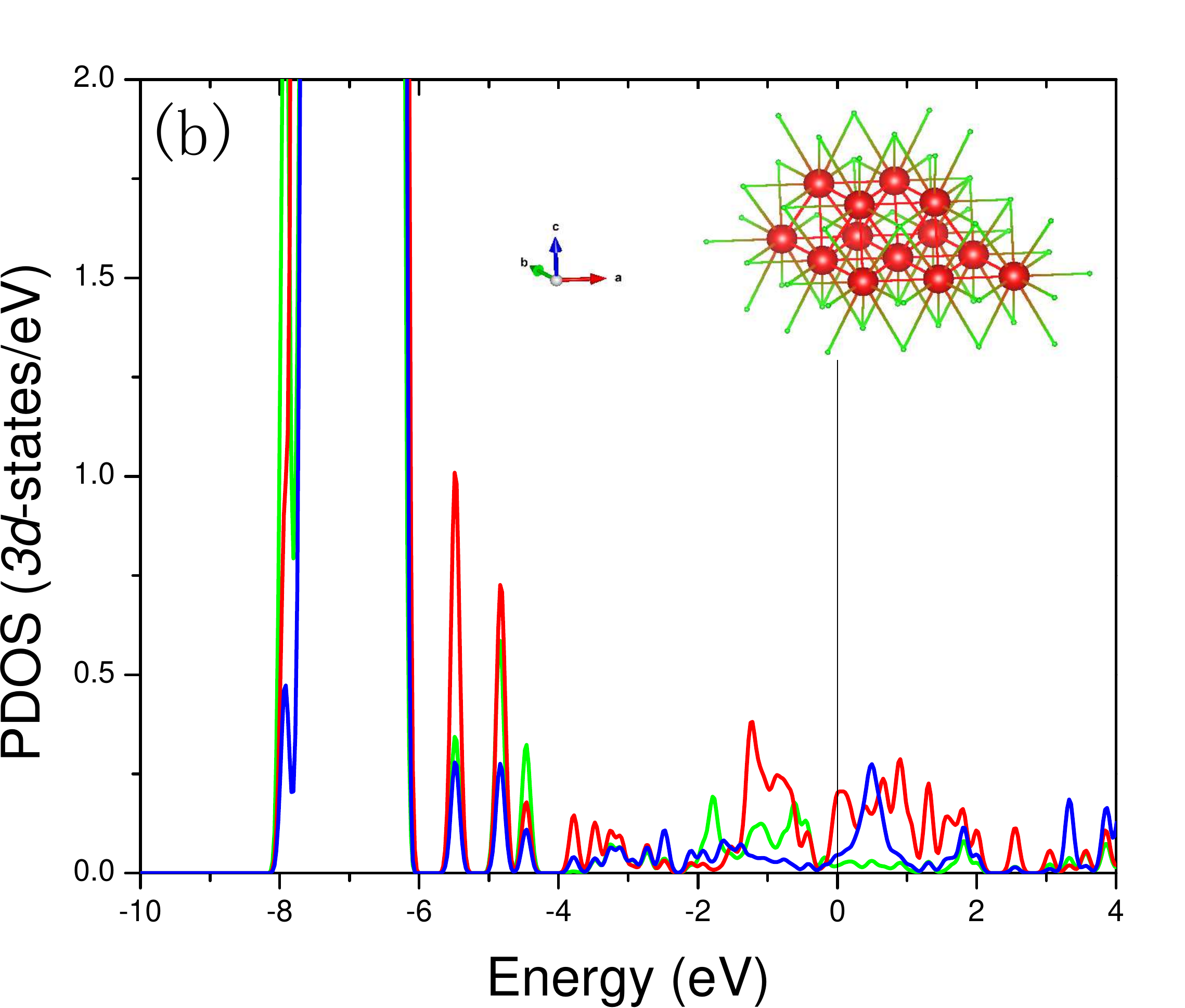}
\caption{(Color online) 
Same as figure~\ref{dos-bulk} but for Zn QDs.
The inset in the right-hand side panel depicts 
a schematic representation of the 
slab used in our calculations, 
arranged in the Zn hcp phase, 
which consist of 13 Zn atoms and 48 H atoms
(red and green symbols, respectively).
}
\label{dos-nano}
\end{figure}

Figure~\ref{dos-nano} shows the calculated PDOS 
for the hydrogen-passivated Zn quantum dot case. 
For the different peaks associated with the 
energies found, both below and above $E_{\text F}$, 
the PDOS shows the structure predicted for 
a box-like quantum dot~\cite{Arakawa,Batabyal-PhyE64,SJLee-SSTech7,Nilius},
namely, a delta function behavior; i.e., 
our results show the effects of quantum confinement~\cite{Harman-qd}.

Furthermore, 
as a consequence of the atomic electronic configuration used in our 
calculations  and due to the strong coupling between the atoms that 
form the QD, the obtained PDOS clearly shows  the
hybridization of the Zn--$4s$, $3p$,  and $3d$ atomic orbitals.

A detailed analysis of the calculated PDOS shows that 
although the Zn--$3d$ orbitals are mainly localized
in the energy range from $-8.2$ eV to $-6.2$ eV, 
non-negligible contributions
of these orbitals to the PDOS over the entire 
energy range shown in the plot are also found. 
\begin{figure}[!t]
\begin{center}
\includegraphics[width=0.65\textwidth]{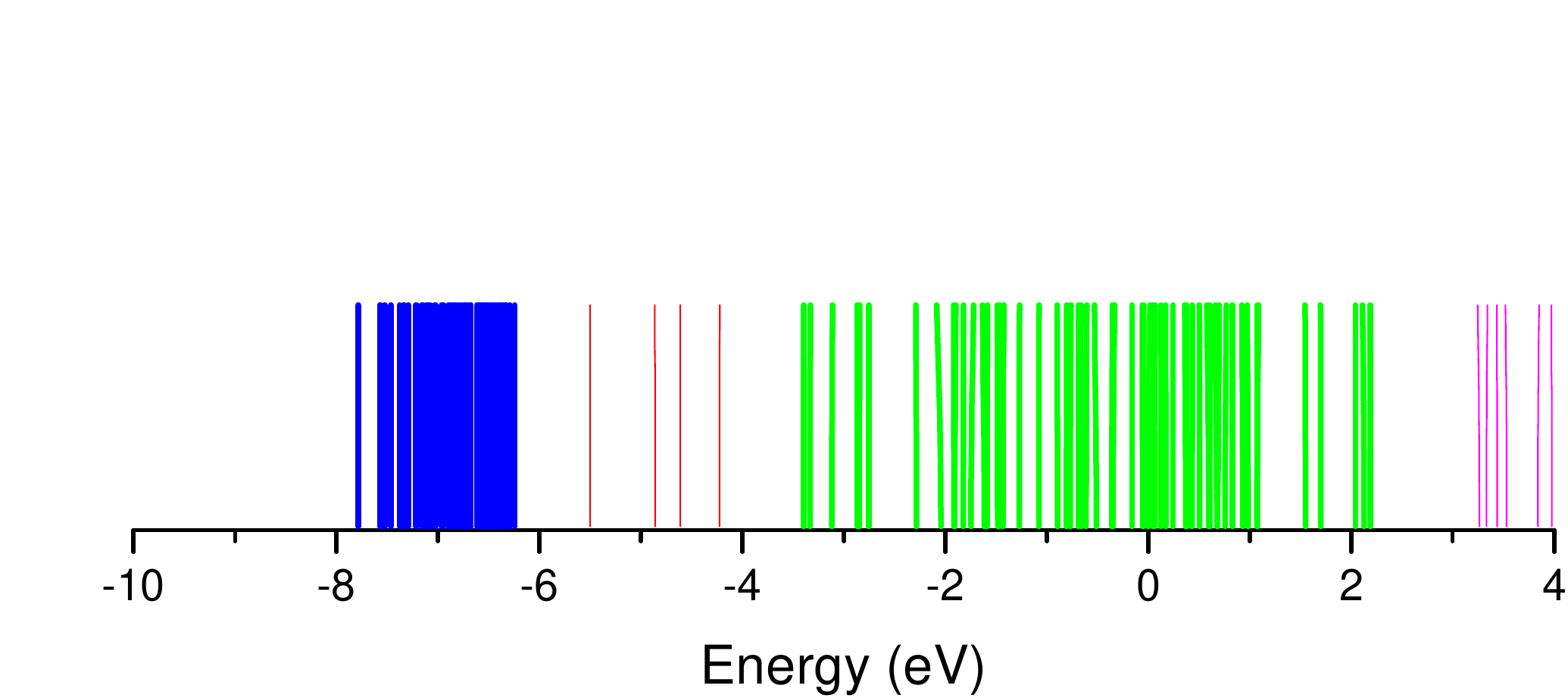}
\end{center}
\caption{(Color online) 
The calculated electronic bands for hcp Zn QDs. 
Colors are a guide to the eye.
}
\label{bnds-nano}
\end{figure}

At the same time, 
for the different energies shown,
important contributions from the $4s$ states (red lines) 
and $3p$ states (blue and green lines)
are also obtained,  
showing the above-mentioned hybridization of the 
Zn--$4s$, $3p$, and $3d$ orbitals.

As in the previous cases, we can use the electronic band structure to 
complement the analysis of the calculated PDOS.
The calculated electronic band structure
for the QD case, 
depicted in figure~\ref{bnds-nano}, 
shows discrete energy levels,
which correspond to the atom-like PDOS shown in figure~\ref{dos-nano}.
From the plot, it is easy to see 
how the energy levels are grouped. 
The lower energy states,
located in the energy range from $-7.8$ to $-6.2$ eV, 
correspond to the $3d$ orbitals that hybridize with 
the $4s$ and $3p$ orbitals. 
In the energy range from $-5.5$ to $-4.2$ eV,
we obtain an isolated set of energy levels
that correspond mainly to the $4s$  and $3p$ states found 
in the PDOS.
In the energy range 
from $-3.4$ to 2.2 eV,
we obtain a group of $4s$  and $3p$ states 
forming a dense set of energy levels,  
and the final energy range, 
from 3.0 eV to 4.0 eV,
shows the last set of energy levels,
whose main contribution arises from the 
$3p$ orbitals.

Although it is found that low-dimensional zinc oxides, zinc chalcogenides,
and other zinc-based systems  have been more studied than the low dimensional Zn 
systems~\cite{CFGuo-ZnO,Azpiroz-ZnS,Kumar-CdS-ZnS,Son-ZnO-graphene,Queiroz,XFang}, considering the lack of information found for 
the layered systems and
metallic quantum dots, we hope that our study 
will be useful and motivate future work. 

At the same time, to avoid speculations and 
compare our results with new and future  data, 
the same for the layered systems as well as for the quantum dot,
more work will be necessary in order to study the 
convergence of the total energy 
as a function of the quantum dot size,
the hydrogen passivation, 
surface effects, 
and other 
model parameters of the  
supercell method used.

\section{Conclusions}
\label{concluye}

We have studied the electronic properties of 
the metallic hcp phase of Zn in the bulk and
low-dimensional 
monolayer, bilayer, and quantum dot 
systems 
from first-principles calculations.
For the bulk case, our results reproduce the previously published 
data on the structural 
and electronic properties.
For the monolayer and bilayer cases, our results 
show that the calculated systems produce lamellar systems,
an interesting property of 
novel 2D systems that  attracts
attention of the scientific community.
By contrast, for the quantum dot system, we found that our calculated
DOS reproduces
the predicted atom-like behavior of 0D systems.
Our calculated cohesive energy for the studied systems, namely 0.73~eV/atom for the bilayer case, 0.56 eV/atom for the monolayer case, 
and 0.29 eV/atom for the QD, show that
although our calculated values are smaller 
the systems are stable.
Considering the interest in and prospects for application of the
different layered systems as well QDs, 
we hope that our work motivates future 
research on these and other non-traditional 
metallic low-dimensional systems.

\section*{Acknowledgements}
Computational resources from 
ABACUS-Cinvestav, Conacyt grant EDOMEX-2011-COI-165873,
and Xiuhcoatl-Cinvestav 
are gratefully acknowledged.

%

\newpage
\ukrainianpart

\title
{Обчислення повної енергії для металічної гексагональної щільноукладеної фази  Zn в об'ємі, багатошарових границях і в границі  квантової точки 
}
\author
 {Д. Олгуін}

\address{
	Центр досліджень та підвищення кваліфікації Національного політехнічного інституту --- підрозділ Керетаро, м. Сантьяго де Керетаро, Мексика
}
\makeukrtitle

\begin{abstract}
\tolerance=3000%
 Досліджено структурні та електронні властивості  металічної гексагональної щільноукладеної фази  Zn в об'ємі, багатошарових границях і в границі  квантової точки, використовуючи обчислення повної енергії. 
З  обчислень густини станів і електронної зонної структури, що узгоджуються з нашими попередніми результатами, отримано об'ємну гібридизацію  $4s$, $3p$ і $3d$ орбіталей  Zn. Крім того, ми знайшли, що гібридизація орбіталей також отримується для моношарових, двошарових систем і систем з квантовою точкою.
Одночасно встановлено, що моношарові і двошарові системи Zn виявляють електронні властивості, характерні для ламеларних систем, в той час, як система квантових точок демонструє поведінку, передбачену для 0D системи.
\keywords обчислення з перших принципів, металічний Zn, квантові точки, 2D системи
\end{abstract}

\end{document}